\documentclass[journal]{vgtc}                
\ifpdf
  \pdfoutput=1\relax                   
  \usepackage{graphicx}                
  \DeclareGraphicsExtensions{.pdf,.png,.jpg,.jpeg} 
\else
  \usepackage{graphicx}                
  \DeclareGraphicsExtensions{.eps}     
\fi%

\graphicspath{{figures/}{pictures/}{images/}{./}} 

\usepackage{microtype}                 
\PassOptionsToPackage{warn}{textcomp}  
\usepackage{textcomp}                  
\usepackage{mathptmx}                  
\usepackage{times}                     
\usepackage{cite}                      
\usepackage{tabu}                      
\usepackage{booktabs}                  

\usepackage{amsmath}
\usepackage{algorithm}
\usepackage[noend]{algpseudocode}

\usepackage[dvipsnames]{xcolor}
\usepackage{xspace}
\usepackage{paralist}
\usepackage{enumitem}
\usepackage{spverbatim}
\usepackage{multirow}
\usepackage[utf8]{inputenc}
\usepackage[english]{babel}
\usepackage{amsthm}
\usepackage{amssymb}
\usepackage{tikz}

\usepackage{graphicx}
\usepackage{wrapfig}
\usepackage{lipsum}  
\usepackage{adjustbox}
\usepackage{flushend}
\usepackage{subfigure}

\theoremstyle{definition}

\newcommand{\eg}{{\it e.g.,\ }}
\newcommand{\etal}{{\it et al.\ }}

\newcommand{\ie}{{\it i.e.,\ }}

\newcommand{\wy}[1]{{\color{Violet} #1}}
\newcommand{\sherry}[1]{{\color{Violet} #1}}

\definecolor{cop}{HTML}{3B4C73}
\definecolor{cobj}{HTML}{036B08}
\definecolor{cpar}{HTML}{0000B3}
\definecolor{cutt}{HTML}{696969}
\definecolor{bluecrayola}{rgb}{0.12,0.46,1.0}

\newcommand{\argu}[1]{\texttt{\emph{#1}}}

\newcommand{\sop}[1]{{\color{cop} \texttt{#1}}}
\newcommand{\sobj}[1]{{\color{cobj} \texttt{#1}}}
\newcommand{\spar}[1]{{\color{cpar} \texttt{#1}}}

\newcommand{\sact}[3]{\texttt{\{\sop{#1}, \sobj{#2}, \spar{#3}\}}}
\newcommand{\sutt}[1]{{\color{cutt} ``#1''}}
\newcommand{\revise}[1]{{\color{black} {#1}}}

\onlineid{1260}

\vgtccategory{Representations {\&} Interaction}

\title{Towards Natural Language-Based Visualization Authoring}

\author{Yun Wang, Zhitao Hou, Leixian Shen, Tongshuang Wu, Jiaqi Wang, \\He Huang, Haidong Zhang, and Dongmei Zhang}
\authorfooter{
\item
 Y. Wang, Z. Hou, H. Huang, H. Zhang, D. Zhang are with Microsoft Research Asia (MSRA).
 E-mail: $\{wangyun, zhith\}$@microsoft.com.
\item
 L. Shen is with Tsinghua University.
 E-mail: slx20@mails.tsinghua.edu.cn.
\item
T. Wu is with Carnegie Mellon University. E-mail: sherryw@cs.cmu.edu.
\item
J. Wang is with Oxford University. 
E-mail: jiaqi.wang@cx.ox.ac.uk.
\item
Work done during L. Shen and J. Wang’s internship at MSRA.
}

\shortauthortitle{Biv \MakeLowercase{\textit{et al.}}: Global Illumination for Fun and Profit}

\abstract{
A key challenge to visualization authoring is the process of getting familiar with the complex user interfaces of authoring tools. Natural Language Interface (NLI) presents promising benefits due to its learnability and usability. 
However, supporting NLIs for authoring tools requires expertise in natural language processing, while existing NLIs are mostly designed for visual analytic workflow.
In this paper, we propose an authoring-oriented NLI pipeline by introducing a structured representation of users’ visualization editing intents, called \emph{editing actions}, based on a formative study and an extensive survey on visualization construction tools.
The editing actions are executable, and thus decouple natural language interpretation and visualization applications as an intermediate layer.
We implement a deep learning-based NL interpreter to translate NL utterances into editing actions. The interpreter is reusable and extensible across authoring tools. The authoring tools only need to map the editing actions into tool-specific operations. 
To illustrate the usages of the NL interpreter, we implement an Excel chart editor and a proof-of-concept authoring tool, VisTalk. We conduct a user study with VisTalk to understand the usage patterns of NL-based authoring systems. Finally, we discuss observations on how users author charts with natural language, as well as implications for future research.

} 

\keywords{Visualization authoring, Natural language interface, Natural language understanding}

\CCScatlist{ 
 \CCScat{K.6.1}{Management of Computing and Information Systems}%
{Project and People Management}{Life Cycle};
 \CCScat{K.7.m}{The Computing Profession}{Miscellaneous}{Ethics}
}

\vgtcinsertpkg

\begin{document}
\maketitle



\section{Introduction}
\label{sec:intro}

Modern visualization authoring systems have emerged to enable creation of expressive visualizations. Nevertheless, they involve complicated GUIs and incur a steep learning curve.
In recent years, as a complementary input modality to traditional WIMP interaction, Natural Language Interfaces (NLI) are adopted to lower the barrier of using advanced visualization tools \cite{Shen2021a}.
In contrast to WIMP interfaces, which require complex menu items and mouse interactions, natural language-based systems require less prior knowledge of user interfaces, and users are not restricted to the locations of menus and buttons to author visualizations~\cite{hoque2017applying, setlur2016eviza}. 

\begin{figure}[!t]
		\setlength{\abovecaptionskip}{1pt}
		\setlength{\belowcaptionskip}{1pt}
	\centering  
	\subfigure[Pipeline of analysis-oriented NLI]{
		\includegraphics[width=\columnwidth]{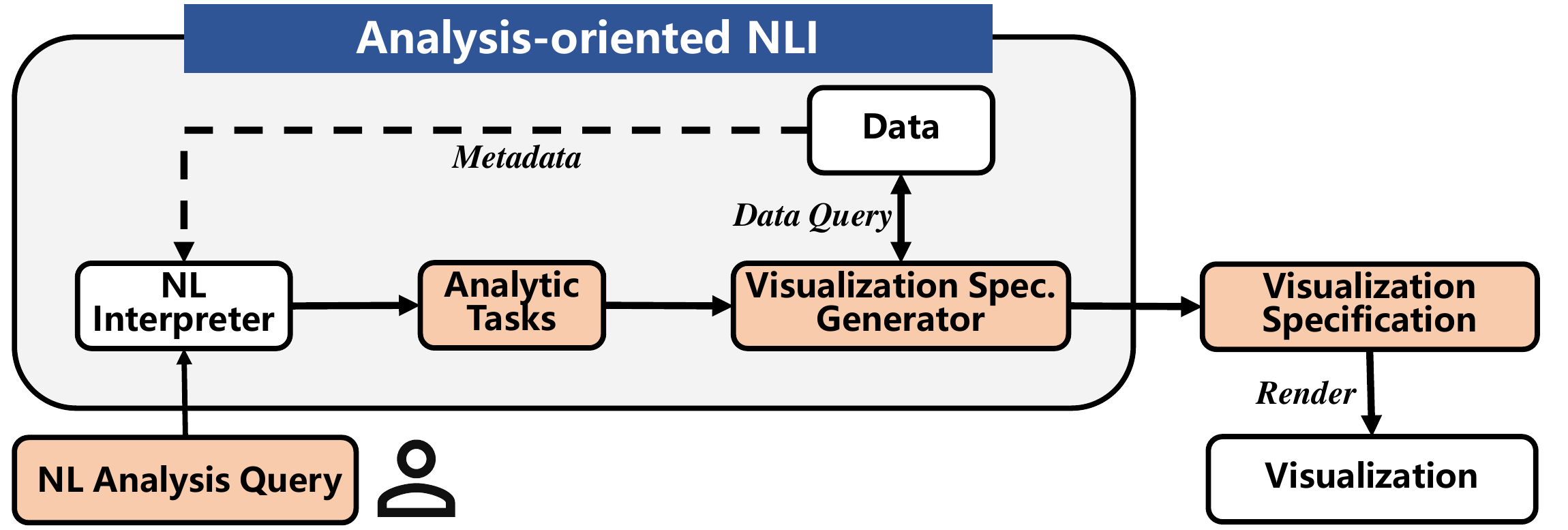}
			\label{fig:analysis_pipeline}
		}
	
	\subfigure[Pipeline of authoring-oriented NLI]{
		\includegraphics[width=\columnwidth]{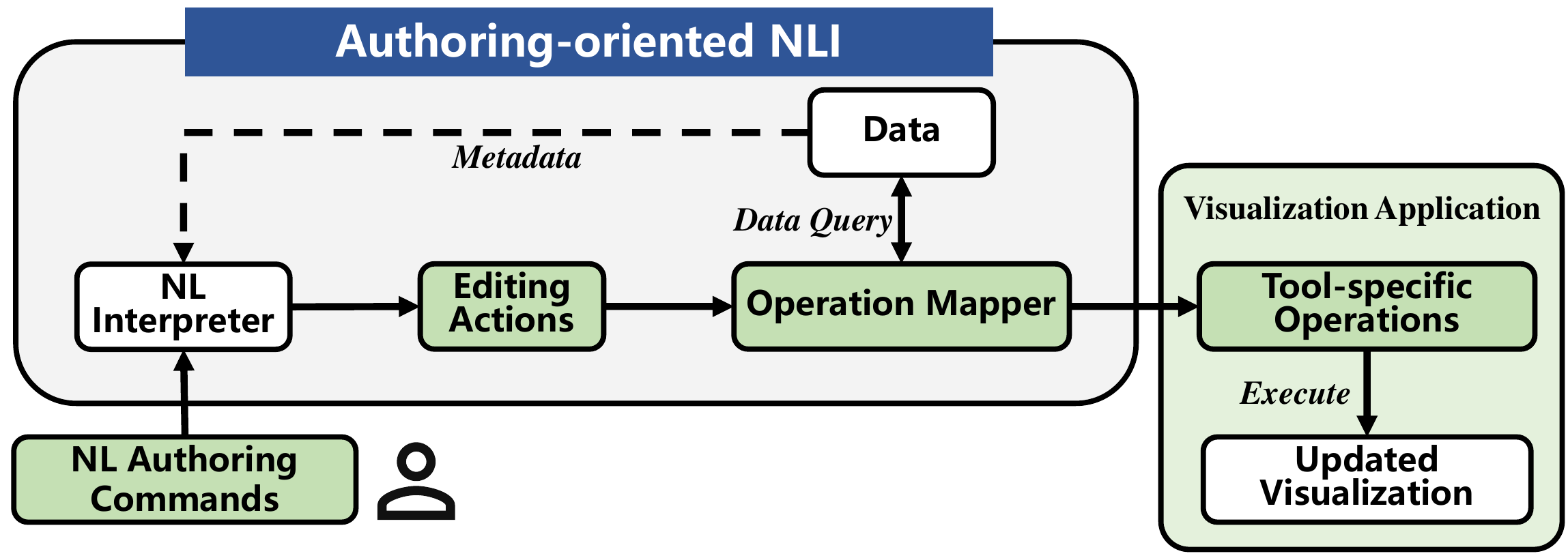}
	\label{fig:authoring_pipeline}
	}
	\caption{Analysis-oriented NLIs generate one-shot visualizations to satisfy users’ analytic tasks, covering subsets of editing operations for visualization authoring. In our authoring-oriented NLI pipeline, we model users' editing intents as \textit{editing actions}, which decouple natural language interpreter and visualization applications as an intermediate layer.}
	\label{fig:pipeline}
	\vspace{-0.4cm}
\end{figure}

While there has been active research into natural language interfaces for visualization systems~\cite{dhamdhere2017analyza, gao2015datatone, setlur2016eviza, yu2019flowsense, srinivasan2017natural}, these systems are primarily designed for analyzing and exploring data.
As shown in Figure~\ref{fig:analysis_pipeline}, an analysis-oriented NLI parses NL queries (\eg \sutt{find the relationship between player goals and salaries across player foot.}) into analytic tasks and data attributes, which are then translated into visualization specifications according to visual design constraints~\cite{narechania2020nl4dv,Shen2021a}.
These specifications may meet the intended analysis~\cite{Shen2021,Amar2005b}, but they may not necessarily be the most \emph{preferred} ones.
In fact, users typically need to change the underlying data, specify visual encodings, and adjust visual presentations like axes, legends, marks, and layouts (\eg \sutt{move the legend to the right of the chart}, \sutt{set mark to woman icon}, and \sutt{change color to pink}). 
Various modern visualization tools~\cite{bostock2011d3, satyanarayan2016vega} and authoring systems~\cite{Kim_2017_7536218, satyanarayan2014lyra, xia2018dataink, Wang:2018:IEC:3173574.3173909, Liu:2018:DIA:3173574.3173697, Ren_2019_8440827} have recognized the importance of rich and flexible data binding and visual configurations; however, analysis-oriented NLIs do not fully support these diverse editing intents.

In this paper, we aim to lower the barrier to supporting NLI in visualization authoring tools. We design a pipeline (Figure~\ref{fig:authoring_pipeline}) that decouples the natural language understanding (with \emph{a natural language interpreter}) and visualization editing command execution (by a \emph{visualization application}).
At the core of the pipeline is a set of \emph{editing actions}. 
These actions are machine-executable commands for modeling the aforementioned visualization editing intents.
They bridge users and visualization applications: the natural language interpreter parses users' utterances into a sequence of such editing actions, and the actions are mapped into tool-specific operations. 
The visualization applications can then adapt and execute the operations to update the visualization. 

We make our pipeline realistic by designing two primary building blocks. First, the formalization of \emph{editing actions}. The role of editing actions in our authoring-oriented NLI pipeline is obvious; however, there is still a lack of comprehensive guidance on how to model visualization editing intents.
To fill in the gap, we conduct a formative study and literature surveys on visualization construction tools and explicate editing actions as mappings between a series of well-defined \textit{editing operations}, their \textit{target objects}, and the corresponding \textit{parameters}.
Second, a \emph{multi-stage NL interpreter}, for parsing users' NL queries into editing actions. It first recognizes data entities and replace them with abstract arguments; then, it uses a deep sequence-labeling model to extract intents and entities from the abstracted utterances; finally, it synthesizes the extracted information into a sequence of editing actions.
With the deep-learning model, we envision that the interpreter can be easily extended to cover wider range of editing actions as we collect a larger corpus of editing utterances. 

Both the editing actions and the NL interpreter can be reused across multiple visualization applications, eliminating the needs to re-design NL interfaces for each one.
Authoring tool developers only need to build an operation mapper to map the editing actions sequence into tool-specific operations.
The only step required for authoring tool developers is to build an operation mapper for mapping editing actions sequences to tool-specific operations.
We enumerate key desiderata in building these mappers, and provide empirical guidance on how to best reuse the provided two building blocks.

We demonstrate the utility of our design with two example applications, \ie an Excel chart editor and a proof-of-concept authoring tool, VisTalk. 
We further conduct an exploratory user study, where participants complete chart reconstruction tasks with natural utterances. Our observations shed light on future work: we should consider how the interpreters can cope with users with different backgrounds, and how the applications can resolve ambiguities and make recommendations when users’ initial command is not supported. 

The contributions of this work are summarized as follows:

\begin{compactitem}
    \item 
    We propose an authoring-oriented NLI pipeline by formulating users' visualization editing intents as \textit{editing actions}, which are atomic units executable by applications. 
    \item 
    We develop a multi-stage natural language interpreter to parse NL utterances into a sequence of editing actions. 
    The interpreter can be reused across visualization applications.
    \item 
    We design an Excel chart editor and a proof-of-concept authoring tool, VisTalk, to demonstrate the utility of the NL interpreter.
    We further conduct a user study with VisTalk to understand how users can construct and edit data charts through natural language. 
\end{compactitem}

\section{Related Work}
Our work draws upon prior efforts in visualization creation tools and natural language interfaces for data visualization.
\subsection{Visualization Creation Tools}
A variety of visual authoring approaches have been proposed to facilitate visualization creation. For example, Polaris~\cite{Stolte2000PolarisAS}, Tableau~\cite{tableau}, and Many Eyes~\cite{ibm} help users conduct data encoding with visual channels in a visualization. These tools enable users to create visualizations in a short time, but they are less flexible. The resulting visualizations are usually standard visual charts. More advanced techniques, such as Lyra~\cite{satyanarayan2014lyra} and iVisDesigner~\cite{6876042}, have been proposed to enable more expressive visualization designs. With these systems, users can easily specify properties of graphical marks and bind them with data. However, these systems only support the modifications of properties by adjusting a set of style parameters. Later, Data Illustrator~\cite{Liu:2018:DIA:3173574.3173697} uses repetition and partition operators for multiplying marks and generating data-driven expressive visualizations. Charticulator~\cite{Ren_2019_8440827} also adopts a bottom-up approach to bind data fields to vector graphics and build visualizations with an emphasis on visualization layouts.
Recently, Data-Driven Guides~\cite{Kim_2017_7536218} has allowed users to draw and design graphical shapes to achieve creative designs. Similarly, DataInk~\cite{xia2018dataink} and InfoNice~\cite{Wang:2018:IEC:3173574.3173909} also bind graphical elements with data fields to facilitate the generation of creative visualizations.

Researchers have explored the design of visualization authoring systems and supported creation from different perspectives. To support visualization creation, the systems have different internal logics.
With these systems, we summarize and categorize most common functions used. We further formulate the functions as editing actions, which is an abstract layer that bridges human intents and editing functions. The editing actions can be explained and executed by the functions supported by different tools.



\subsection{Natural Language Interfaces for Visualization}
Natural language interfaces have been adopted extensively to improve the usability of visualization systems\cite{Shen2021a}. 
Commercial tools like IBM Watson Analytics~\cite{ibm}, Microsoft Power BI~\cite{pbi}, Tableau~\cite{tableau}, ThoughtSpot~\cite{ThoughtSpot}, and Google Spreadsheet~\cite{dhamdhere2017analyza} automatically translate the natural language questions to data queries and present query results with visualizations. However, these systems limit natural language interactions to data queries and corresponding standard charts.

Extensive research has been devoted to a better experience of using natural language for visual analytics~\cite{Shen2021a}. 
To complete the analytical tasks, they treat user input utterances as natural language queries, and translate them to logic query languages~\cite{androutsopoulos1995natural,li2007nalix,zhong2017seq2sql}. Researchers further study the interactive visual analytic systems that support natural language queries.
Wen \etal~\cite{wen2005optimization} design a system to update the visualizations based on user queries dynamically. 
Articulate~\cite{aurisano2016articulate2} generates a graph to select proper visualizations to answer users' natural language questions and complete analytic tasks.
ADVISor\cite{Liu2021b} and ncNet\cite{Luo2021a} leverage deep learning models to translate an NL query to a visualization.
Going beyond data exploration,
FlowSense~\cite{yu2019flowsense} applies natural language techniques to dataflow visualization systems.
Gridbook\cite{Wanga} eases the difficulty of writing formulas on the spreadsheet grid by supporting formulas expressed in natural language.
Vis-Annotator\cite{Lai2020a} accepts textual descriptions and outputs annotated visualizations while Kim \etal\cite{Kim2020a} focus on answering questions about the given visualization.  
Srinivasan \etal integrate natural language interfaces to facilitate multimodal interaction for visual exploration\cite{srinivasan2017orko,srinivasan2020inchorus,Saktheeswaran2020,srinivasan2020interweaving,srinivasan2019discovering}.

Researchers also try to address ambiguity problems through different means. On the one hand, researchers attempt to address ambiguities through the design of interactive systems to invite users to clarify their requests. For example, DataTone \cite{gao2015datatone} introduces interactive widgets to address ambiguity problems.
Eviza \cite{setlur2016eviza} further enhances interactions by providing graphical answers that can be directly manipulated. 
Iris~\cite{fast2018iris} is a conversational interface that enables users to combine commands through many rounds of question-answering.
Sneak pique\cite{Setlur2020} explores autocompletion to help users formulate analytical questions while Snowy\cite{Srinivasana} recommends utterances for conversational visual analysis.
On the other hand, design principles are also summarized to improve the understanding of users' utterances.
To enhance natural language interactions, Hoque \etal \cite{hoque2017applying} apply pragmatics principles and propose a theoretical framework for interaction with visual analytics.
Hearst \etal \cite{hearst2019toward} conduct an empirical study to explore default visualizations for vague expressions in natural language queries. Similarly, Setlur \etal \cite{setlur2019inferencing} explore inferring underspecified natural language queries and propose a systematic approach to resolve partial utterances.

Existing work enables the natural language interfaces in a case-by-case manner, and mainly focuses on visual analysis. 
In comparison, we formalize users' diverse visualization editing intents into \textit{editing actions}, covering a richer set of flexible visual configurations.
The editing actions bridge the gap between the NL interpreter and visualization applications.
The design and implementation of the natural language interpreter can be reused across applications, alleviating the burden of re-designing NLI per application.

\section{Formative Study}
User utterances are apt to be free and variant among usage scenarios. 
To better understand how visualizations are created with natural language, we conduct a formative study in the form of one-hour meetings with six participants on how they express their intentions for visualization authoring. Understanding usage patterns is conducive to the design of our natural language interpreter. 
 All the participants had previously created visualizations for presentation purposes (e.g., data reports, presentation slides) in their daily work. They have used GUI-based (e.g., Excel and Power BI) or code-based (e.g., Python) tools to create visualizations. All of them have conducted configurations to standard visual charts or created expressive visualizations such as pictographs. Two create charts for data analysis as their daily work. Three other participants have used design tools (e.g., Adobe PhotoShop and Illustrator).

\subsection{Procedure}
The meeting was started with a semi-structured interview. During the interview, the participants were asked to describe their process when developing visualizations, as well as the major challenges they face when they want to ask a system to do the job for them. Then, they were asked to list example natural language sentences to describe the operations and commands they will give if an intelligent assistant can do the job for them. 
Then, we collected six visualizations by sampling charts from systems like Excel, Tableau and more expressive chart designs from InfoNice \cite{Wang:2018:IEC:3173574.3173909} and Charticulator \cite{Ren_2019_8440827}. We told the participants to imagine using an authoring system that would correctly understand all the commands and update the visualization step by step. We asked them to articulate a list of natural language commands that they would use to clearly describe the steps to create these visualizations. These exercises helped us understand our users' natural way of thinking and workflow through concrete examples.

\subsection{Results}
All of our participants expressed interests in having a natural language-based tools. They felt a natural language-based authoring tool is useful especially when they are not familiar with the user interfaces, which are more likely to happen when they need to create bespoke visualizations and involves a large number of steps in design tools. Here we summarize the common design practices.

\textbf{General Workflow.}
All of the participants showed similar preferences of the general workflow of creating visualizations with natural language. In general, they preferred a top-down method for basic charts such as bar chart, column chart, pie chart, and so on. More specifically, they usually started with the chart type and encoding, then followed up with more configurations. The names of the charts give them the convenience of defining data encoding and basic properties of a visualization with several simple words, such as ``Show me the sales by year with a line chart.'' Based on the chart generated by the system, users could do more customization. 
At the same time, two of the participants also tried to describe and control the data encoding by themselves. For example, ``Bind sales to y axis and year to x axis.'' They felt it especially useful when they were not sure of the name of the charts to create, or when they thought they were creating a bespoke chart that the system may not support by default. 
As the final step, participants preferred to further fine tune the visualizations by adding or removing chart components, or modifying the default formatting of charts.

\textbf{Intention Description.}
When conducting authoring, users expected the systems to react directly to each of their natural language commands. Sometimes, they used simple sentences, for example, ``change the color of the bars'', or ``please make the title bold''. They might also use complex utterances with more than one intent is included. For example, the sentence ``I want a chart with rounded rectangles, showing me the distribution of the data'' implies users wanted a column or bar chart with rectangles replaced with rounded shapes. Besides, the complexity of users' input may relate to users' experience of using natural language agents and visual chart creation. In our study, the more experienced the user is, the simpler the commands are. This might be because experienced users have lower expectations for parsing capabilities of natural language agents.

\textbf{Visual Property Setting.}
When referring to the objects, they also used data labels. For example, ``make the China bar red.'' Some users prefer to use some properties to refer to an object. For example, they may use ``red line'' or ``dashed line'' to refer to a red reference line on a bar chart. Participants also tried to use pronoun (e.g., it, them) to refer to the objects, especially when they wanted to specify several properties of one visual element successively.
When describing visual properties, our participants felt it easy to describe general properties like ``blue'' for colors, ``$> 200$'' for filters, ``dashed line'' for formatting, etc. They might directly use common adjectives, or the category name for these properties. 
By contrast, it is harder to describe degree, value, extents or positions precisely. As a result, our participants preferred to use relation to describe degree, level, extent, value, position, etc, such as ``on the top of A'', ``larger than B'', ``lower than C'', ``darker than current one'', ``wider than D''. 

The results of our formative study show the diversity of users' expressions when they are not given guidance on their NL input. A structured representation is required to execute users' intents. Further, we need to formulate them into intent units that can be combined so that users' utterances can be represented with flexibility.


\section{Editing Actions}
As illustrated in Figure \ref{fig:pipeline}(b), the core of our pipeline is an intermediate layer called \emph{editing actions}, which are a list of structured descriptions of user intents that can be executed by visualization applications.
We design editing actions to bridge users and applications:
the natural language interpreter parses users' utterances into a sequence of these actions, and the actions can then be adapted and executed by the applications.
On the one hand, these editing actions are somewhat akin to the declarative languages of visualizations (\eg Vega-Lite), in the sense that they offer an implementation-independent bridge between the NL and the visualization. On the other hand, the editing actions differ from existing declarative grammars of visualizations, because they model the \emph{delta} of visualization (\eg \sutt{Turn the dashed lines black and thicker}), and is a potential first step towards authoring-oriented declarative language.
Below, we formalize the model of editing action representation based on the formative study and surveys of prior work, and discuss the necessary designs to address vague, under-specified, or context-dependent descriptions in user utterances.

\begin{figure*}[!t]
		\setlength{\abovecaptionskip}{3pt}
		\setlength{\belowcaptionskip}{3pt}
    \centering
    \includegraphics[width=0.9\linewidth]{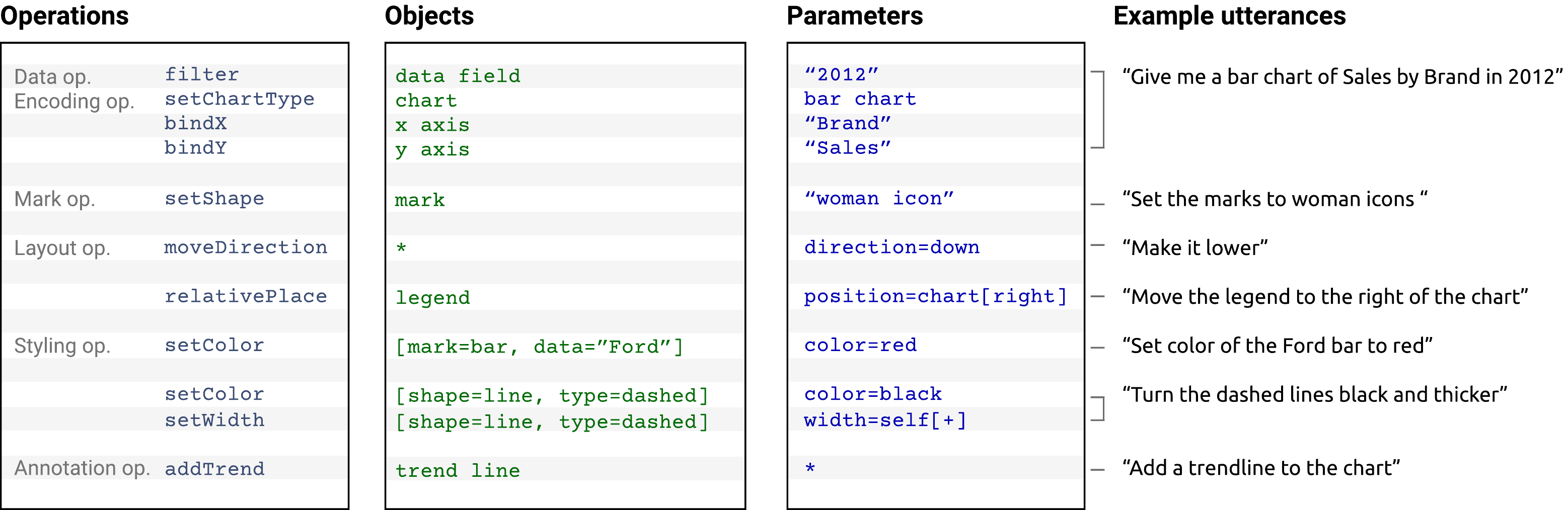}
    \vspace{0px}
    \caption{An utterance can be mapped into a series of editing actions.  An editing action consists of three parts: operation, objects, and parameters.
    The examples show the usages of object selector, vague parameters, and relational parameters. * denotes unknown properties.}
    \label{fig:taxonomy}
     \vspace{-12px}
\end{figure*}

\subsection{Editing Action Representation}\label{sec:actions}

Various systems have covered a wide range of functions to enable the composition of both standard charts and highly customized interactive visualizations. 
However, there was no common design model that can let us consistently describe visualization editing intents~\cite{satyanarayan2019critical}.
\revise{To address the lack of modeling, we survey and unify intent descriptions from our formative study and modern visualization construction systems, which all involve gradual editing and refinement.
In particular, following the visualization guidelines\cite{Sawicki2022,Kandogan2016,Diehl2020}, we examine editing actions involved in creating and editing visualization designs in commercial products or online demos\footnote{The editing actions are collected from the tools' official introduction websites and third-party online tutorials.}, including Microsoft Excel, Tableau, Voyager \cite{Wongsuphasawat2017Voyager2A}, Data Illustrator~\cite{Liu:2018:DIA:3173574.3173697}, ChartAccent~\cite{ren2017chartaccent}, Charticulator~\cite{Ren_2019_8440827}, and InfoNice~\cite{Wang:2018:IEC:3173574.3173909}.


\revise{These systems widely cover the three common categories of visualization construction tools~\cite{satyanarayan2019critical}, namely, template editor, shelf-construction, and visual builders.}
Each of the categories construct charts with different editing workflows: 
(1) In \textit{template editor} tools like Microsoft Excel and RAWGraphs\cite{Mauri2017}, users frequently start with high-level command queries to generate template charts before making further extensive customization.
These commands omit details on visualization channels, and instead stress chart types and targeted data entries.
For example, to query a bar chart with y-axis encoded with Sales and x-axis encoded with different countries from these tools, the most natural driving utterance might be \sutt{Give me a bar chart for \textit{Sales} by \textit{Country}}.
(2) Tools supporting \emph{shelf-construction} (e.g., Tableau, Voyager~\cite{Wongsuphasawat2017Voyager2A}) builds charts by mapping data fields to encoding channels (e.g., x, y, color, shape). 
To construct the same example above, the utterance that best aligns with these tools would be \sutt{Bind \textit{Sales} to y-axis} and \sutt{Bind \textit{Country} to x-axis}. 
(3) \emph{Visual builders} tools like Data Illustrator \cite{Liu:2018:DIA:3173574.3173697}, Charticulator \cite{Ren_2019_8440827}, ChartAccent \cite{ren2017chartaccent}, and InfoNice \cite{Wang:2018:IEC:3173574.3173909}
would gradually tune marks, glyphs, coordinate systems, and layouts. 
Users might start with \sutt{draw a rectangle}, and then, \sutt{repeat it horizontally on \textit{Country}}, and finally \sutt{bind height to \textit{Sales}} to arrive at a similar bar chart. 

\revise{
Comparing across the aforementioned chart editing actions, we make two observations: (1) The levels of editing intent vary greatly from single element editing to integrated construction; and (2) despite the varying intents, the operation and its targeted  data entries remain present in all utterances.
Therefore, we consider \textit{\color{cop}operations}, \textit{\color{cobj}objects}, and \textit{\color{cpar}parameters} as the core entries, and formally propose the notion of an editing action as a combination of the three:
\begin{equation*}
    \texttt{editing~action := \sact{operation}{objects}{parameters}}
\end{equation*}

The \textit{\color{cop}operation} identifies the type of editing intent (e.g., \sutt{change color}, \sutt{add annotation}).
With a single notion of ``operation'', we maintain flexibility across varying levels of editing intent (e.g., low-level operations like \sutt{Bind \textit{Sales} to y-axis} versus high-level ones like \sutt{Give me a bar chart for \textit{Sales} by \textit{Country}}), and therefore can fit into different styles of visualization tools.
Meanwhile, the \textit{\color{cobj}objects} are the targets that the operation applies to, such as the canvas area, the bars in a bar chart, the title text, etc, whereas the \textit{\color{cpar}parameters} indicate the degree or configurations of the operations.
}
Figure \ref{fig:taxonomy} is an example that illustrates the process to author an annotated bar chart through natural language.
We will describe them in detail below.


\subsubsection{Operations}

We first go through the category of Vega-Lite grammar~\cite{satyanarayan2016vega}, and a large number of editing operations related to \textit{data}, \textit{encoding}, and \textit{mark}.
To support more expressive changes for communication and presentation, we further survey ChartAccent \cite{ren2017chartaccent}, Charticulator \cite{Ren_2019_8440827}, Data Illustrator \cite{Liu:2018:DIA:3173574.3173697}, and InfoNice \cite{Wang:2018:IEC:3173574.3173909} to develop three more categories related to \textit{layout}, \textit{styling}, and \textit{annotation}, which are not covered by analysis-oriented NLIs}.
They form the eventual six main categories, which we use as guidance for designing the concrete operations (examples are listed in Figure~\ref{fig:taxonomy}):
\begin{compactitem}
    \item \textbf{Data operations} are those that retrieve and calculate data from datasets. Typical operations include filter, aggregate, bin, set time unit, and sort.  
    \item \textbf{Encoding operations} are the ones that bind data fields to different encoding channels. Users may specify encodings by explicitly binding elements, or by assigning chart types. 
    \item \textbf{Mark operations} refer to the configurations related to the symbols that encode data in a visualization. 
    By changing the attributes of marks like shape, color, and style, users may implicitly customize the encoding styles applied to all related data points (\eg use circles rather than rectangles in shape encodings).
    \item \textbf{Styling operations} include graphical and textual edits. Graphical operations can change the color, size, shape, icon, and stroke of graphical elements (\eg axis lines), whereas textual operations are acted on text-related properties such as font and content. 
    Compared to mark operations, styling specifically refine on elements without data encoding.
    \item \textbf{Layout operations} concern the positions and offsets of chart elements (\eg annotations and legends). Users can place an object at a designated position, or move it along a certain direction.
    \item \textbf{Annotate operations} manipulate annotations that enhance the charts. Common types of annotations include labels, annotation text, reference line/band, trend line, average line, etc.
    They can be bounded with data; for example, trend lines can show the trend between two time points. 
\end{compactitem}



\subsubsection{Objects}
Objects are the components that operations target. 
For basic visual charts, the objects include marks, axes, titles, legends, gridlines, etc. More advanced objects include annotations such as trend lines, reference lines, reference bands, text annotations, and embellishments.  
These objects correspond to different operations. For example, the objects of \textit{data operations} include data fields, data points, data type, data range, etc. The objects of \textit{encoding operations} include the data fields, chart components (e.g., axes, legends, and title), and mark channels (e.g., mark size, mark shape, and mark position). The objects of \textit{styling operations} include visual properties such as opacity, stroke, text font, text color, etc.

Identifying objects from user utterances can be nontrivial for two reasons.
First, the objects in an editing action can be underspecified. For example, a user may say \sutt{give me a chart} after uploading a dataset. 
We directly model users' intent and use a special notation \texttt{``*''} to represent the under-specified objects, 
such that the explicit reference can be deferred to the visualization system implementation.

\revise{Further, rather than using standard terms like ``mark'', users may refer to objects with descriptive languages on \textit{object properties} (\sutt{red line}, where it is not clear whether the line is a mark or an additional visual shape), or with names created on-the-fly (\eg users may simply name the chart as ``US2020'' after applying a filter ``US'' and ``2020'' to the chart) \cite{hoque2017applying, laput2013pixeltone}. 
We help capture such reference patterns with object selectors and naming:
}

\textbf{Object Selectors.}
We use object selector to process implicit filters.
Consider the intention \sutt{turn the red line blue}.
While it does not directly specify an object, we use the selector to present it as \sobj{[shape=line,color=red]}, \ie to select objects with the properties ``line'' and ``red.''

\textbf{Object Naming.}
We allow dynamically assigning names to chart to reflect the aforementioned use case of ``US2020''.
Besides easy reference, object naming further enables nested designs of charts. For example, users may put two charts together by opening up a new canvas and saying \sutt{put \textit{US2019} and \textit{US2020} side by side.} This also helps enable creative visualization design. 
Figure \ref{fig:examples}(c) shows an example where the user uses two icons to create a pictograph and reuses it in another chart to create novel visualization designs.

\subsubsection{Parameters}\label{subsec:parameter}
Parameters are specific configurations of the operations. 
They correspond to the operations and objects, describing to what extent the operations are performed, or how operations are applied to objects. 
As a result, the parameter types vary with object properties. 
They can be numerical values for chart width and height, enumerations for font style and encoding channels, and boolean values for adding or removing components.
Figure~\ref{fig:taxonomy} shows example parameters corresponding to different chart objects. 
While the parameters can be definite in some cases (\eg the \emph{absolute} pixel \emph{quantity} in \sutt{set the chart 10px wide and 13px height.}), they sometimes take more \emph{qualitative} or \emph{relative} forms:

\textbf{Vague Parameters.}
Utterances usually contain free-form parameters that sound natural to humans, yet are hard-to-decode for the machines \cite{hearst2019toward}. 
We collect these keywords as part of the parameter library, and expect the application tool to further interpret and map them into machine-understandable values based on design environments.
For example, for color values, we support a list of color names (e.g., \spar{color=red}, \spar{color=navy~blue}). We also define a number of extent keywords, ranging from ``extremely'', ``very'', ``moderate'', ``little'', to ``very little''. Users can therefore describe the adjustments of size as ``make it really large'', which will be mapped to ``very large'' and can be parsed as \sact{setSize}{*}{ size=very.large}.

\textbf{Relational Parameters.}
Users also tend to make relative statements anchoring on some existing visual elements, \eg \sutt{make it larger}, or \sutt{set the color of the title darker than the bars}.
For these relational parameters, we expect users to specify the objects it compared to, and the direction of change. For example, \sutt{make the title darker} is explained as make \emph{itself} darker, where the object is the title; \sutt{place the legend on the right of the plot area} is explained as \emph{right} to the \emph{plot}. 
Formally, the utterances are interpreted to \sact{setColor}{title}{color=self[darker]} and 
\sact{place}{legend}{position=plot[right]}. 
\spar{self[darker]} means the color darker than it self while \spar{plot[right]} means on the right of the plot. Applications should further determine the exact values of these parameters after receiving these vague parameters.

\subsection{Example Utterances}\label{sec:example}
Combining operations, objects, and parameters, authoring actions can represent various authoring utterances. An utterance may corresponds to multiple editing actions. 
Here we show example utterances and the corresponding editing actions to further illustrate the use of editing actions. In Table \ref{tab:examples}, the utterances are represent with one to four editing actions. These examples include object selectors (d, e),  vague parameters (b, c, d, e, f), and relational parameters (b, d).

\begin{table}[t]
\centering
\includegraphics[width=0.95\columnwidth]{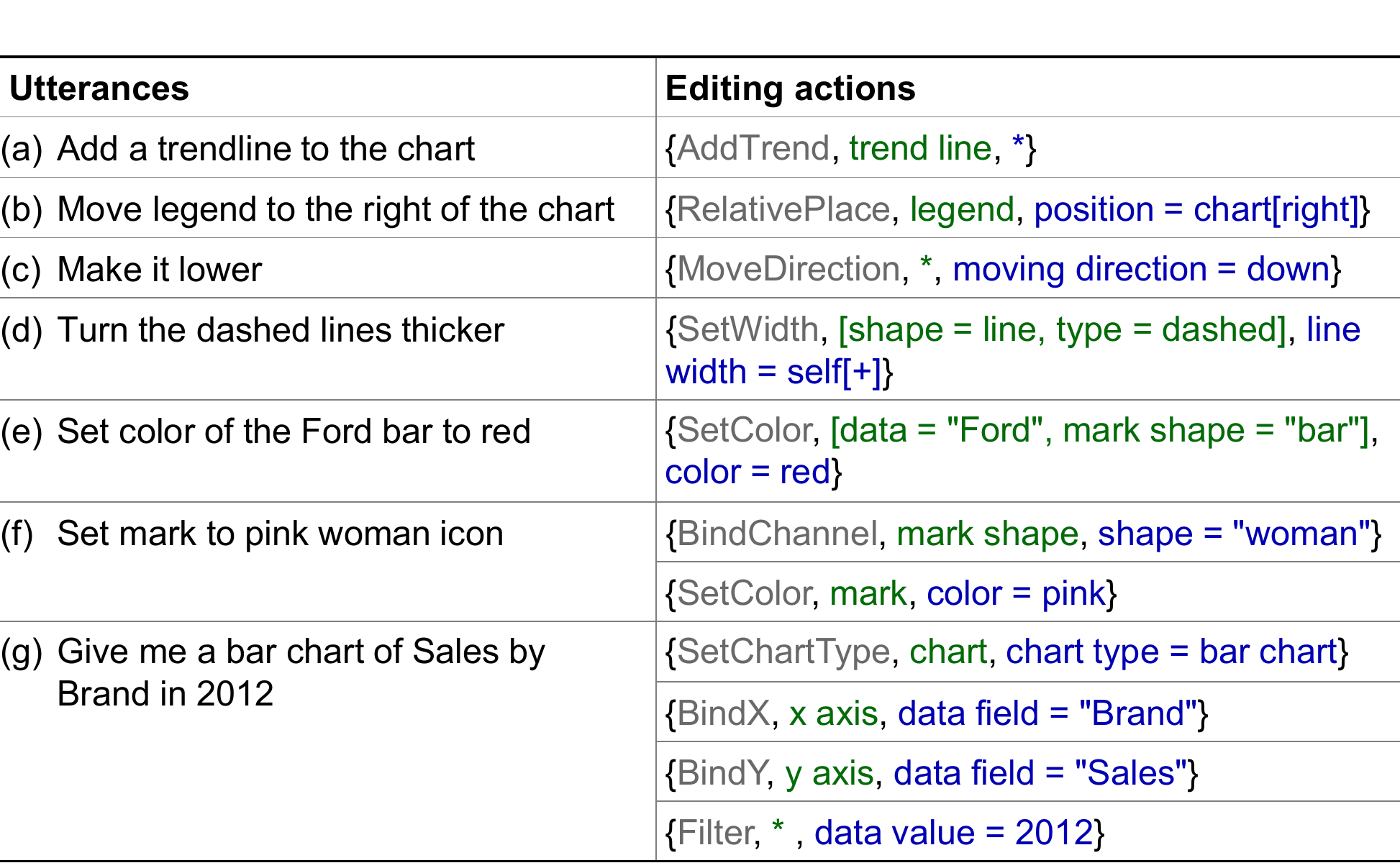}
\vspace{1px}
\caption{Example utterances (left) and corresponding editing actions (right). One utterance may correspond to multiple editing actions.}
\label{tab:examples}
 \vspace{-12px}
\end{table}

\section{NL Interpreter}
\label{sec:interpreter}

\begin{figure}[t]
	\setlength{\abovecaptionskip}{2pt}
		\setlength{\belowcaptionskip}{3pt}
\vspace{1px}
\centering
\includegraphics[width=\columnwidth]{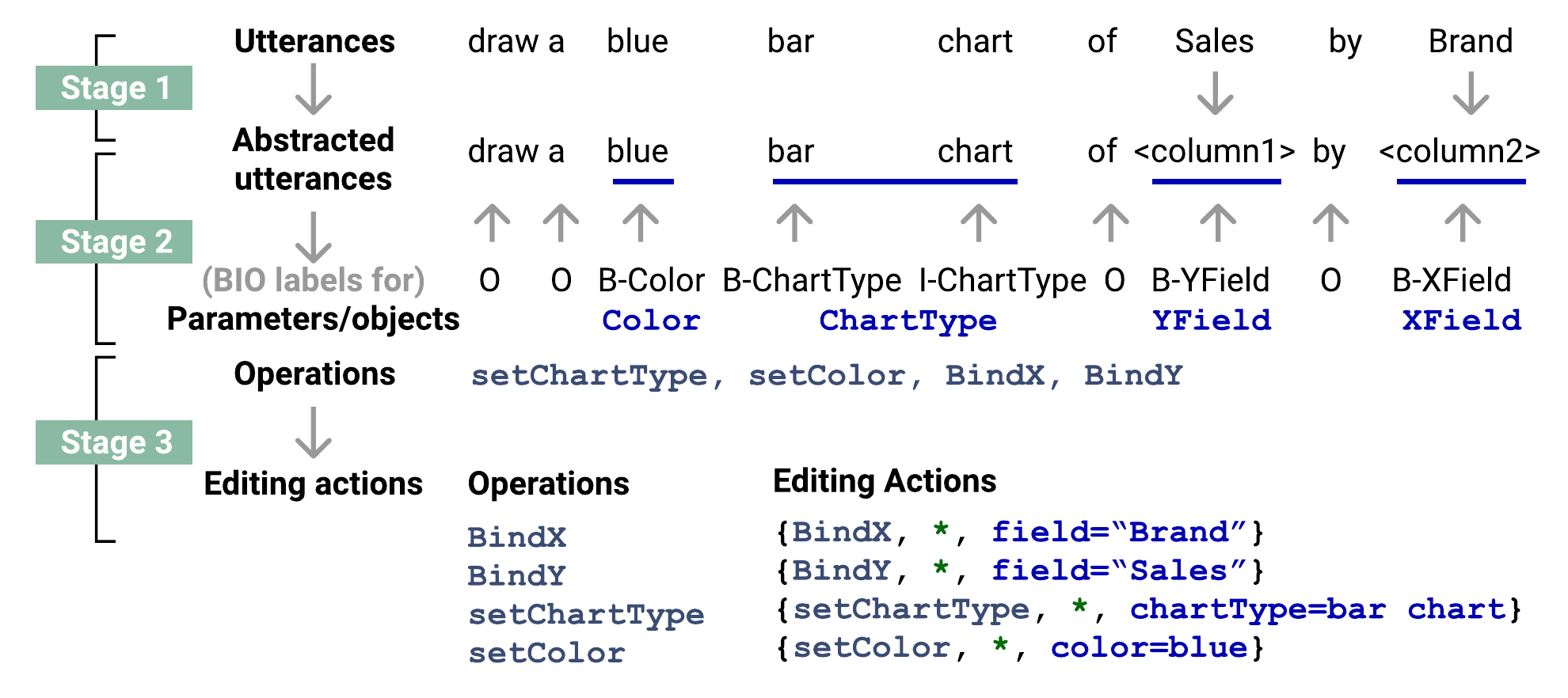}
\vspace{0px}\caption{A multi-stage natural language interpreter: A natural language utterance is first abstracted by replacing data-related entries. Then, we extract operation categories. Following the BIO format\cite{ramshaw1999text}, we can map the words in the sentence into a sequence of labels, where B-, I-, and O represent begin, inside, and outside. The extracted parameters, objects, and operations are further synthesized into editing actions.}
\label{model_example}
 \vspace{-16px}
\end{figure}

\revise{
To demonstrate the efficacy of the editing actions and enable visualization applications to assess our modeling, we provide proof-of-concept design and implementation of an NL interpreter.
}
\revise{For an input utterance, NL interpreter aims to translate it to a sequence of editing actions that will then be passed to visualization applications. As shown in Figure~\ref{model_example}, we design a multi-stage interpreter to parse a natural language utterances into editing actions. First, it identifies user intents in the utterance as operations; then, it extracts useful parts as objects and parameters for the targeting operations; finally, it recognizes the relations between the identified operations, objects and parameters, and organizes them into editing action tuples.}

\subsection{Stage 1: In-context Data Entry Abstraction}\label{sec:abstraction}
To reduce the sparsity and uncertainty of intent recognition, we use a data entry recognizer to identify and abstract dataset-related data entries in the utterance.
The recognizer enumerates through the N-grams in the utterances, checking for the similar entities (words or short phrases) that can be matched with the data attributes, including the \emph{table names}, \emph{column names}, and \emph{cell values} of a dataset table. To resolve ambiguities, we compute the similarity between the tokenized entry and the data attributes in a similar way to NL4DV \cite{narechania2020nl4dv}. 
Once we completed data entry recognition, we replace the entities with placeholders like \argu{<column>}, \argu{<value>} to obtain the \emph{abstracted utterances}. In the case of special numbers and dates, we further specify value into \argu{<integer>}, \argu{<float>}, \argu{<date>}, and \argu{<year>}. 

\subsection{Stage 2: Information Extraction for the Editing Action}
At the core of the interpreter is the ability to parse the utterances into {\color{cop}operations}, {\color{cobj}objects}, and {\color{cpar}parameters}.
Operations are usually summarized or abstracted from utterances, whereas objects and parameters are directly extracted.
As such, while there might be other alternatives, we treat this step as two NLP sub-tasks:
(1) a multi-label classification task, to detect potentially multiple operations from an utterance; and 
(2) a sequence labeling task using BIO sequence tagging~\cite{ramshaw1999text}, to recognize token chunks that represent objects and parameters from one sentence.
Specifically, we chose to frame extraction step as sequence tagging (as oppose to \eg end-to-end NL to code translation) for its richer interpretability. 
With fine-grained semantic annotation on each entity, it is easier for people to inspect and potentially amend the parsing result of a given utterance.

As shown in Stage 2 of Figure~\ref{model_example}, we design a deep-learning model to  simultaneously performs the aforementioned two tasks, outputting separated lists for operations, objects, and parameters.
While various prior work used rule-based methods (\eg both FlowSense~\cite{yu2019flowsense} and NL4DV~\cite{narechania2020nl4dv} use lexical and dependency parsing structures) for precise recognition, we argue that heuristic rules usually only handle limited forms of utterance.
In comparison, deep-learning models are more capable of flexibly interpreting diverse utterances. One can easily extend the capability of the NL interpreter by adding training examples that express intended operations, objects, and parameters~\cite{xu2013convolutional,guo2014joint,zhang2016joint,liu2016attention}.

\begin{figure}[!t]
\centering
\includegraphics[width=0.95\linewidth]{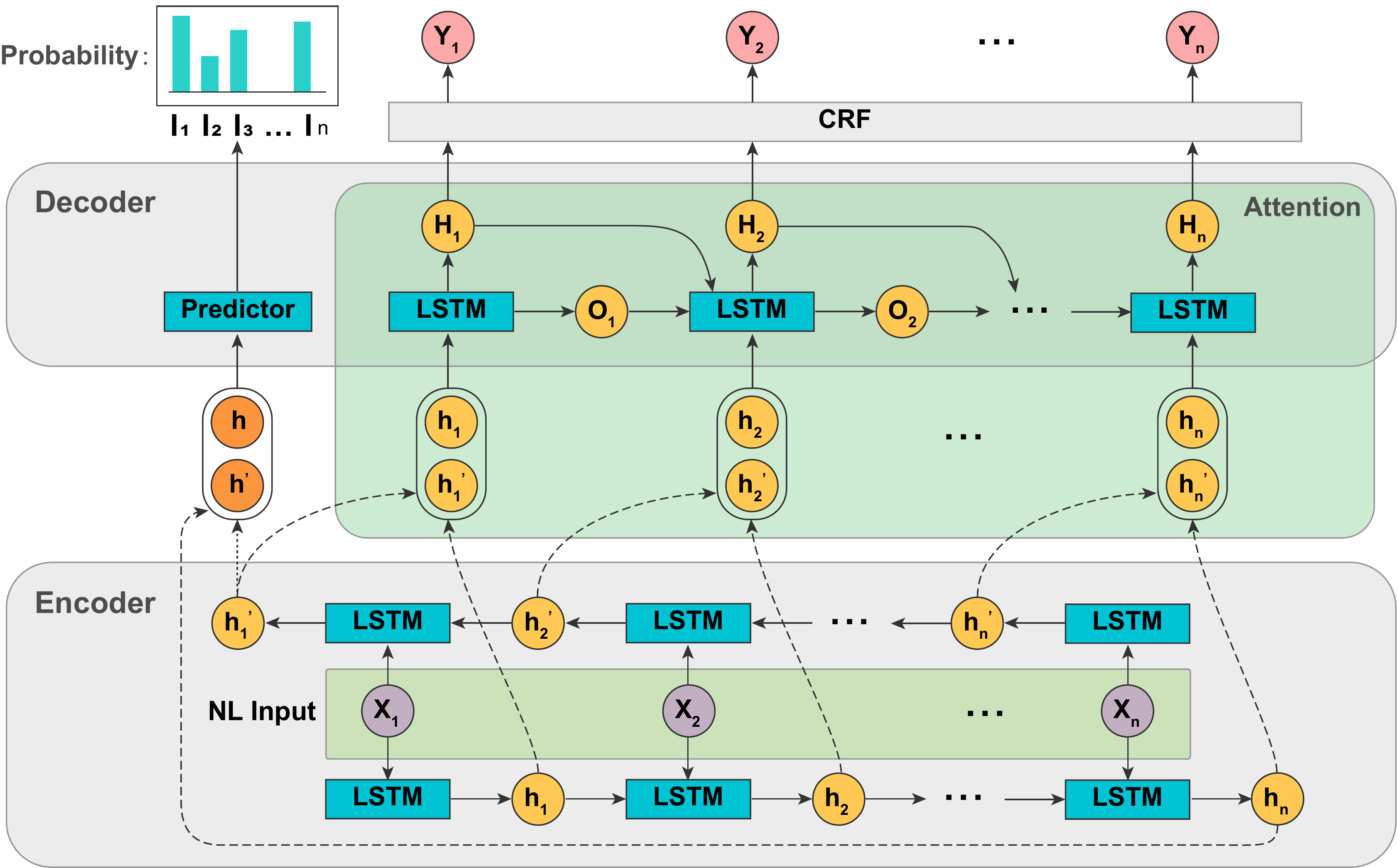}
\caption{The architecture of the deep learning model for intent and entity extraction, which is based on the encoder-decoder framework, attention mechanism, and CRF algorithm.}
\label{model_arch}
 \vspace{-10px}
\end{figure}

In specific, as shown in Figure~\ref{model_arch}, the neural model is based on the encoder-decoder framework~\cite{liu2016attention}, attention mechanism \cite{vaswani2017attention}, and Conditional Random Fields (CRF) algorithm~\cite{lafferty2001conditional}.
We chose the model for its balanced quality and efficiency.
On the one hand, Bi-LSTM-CRF is commonly used for similar sequence tagging tasks like Named Entity Recognition, and empirically we found its accuracy to be sufficiently high;
On the other hand, the model is lightweight enough that it can be deployed and integrated into various different platforms without causing much latency (unlike some pre-trained models, \eg BERT).

We train the model on a dataset containing a mix of real and synthetic utterances expressing certain visualization editing intents.
We first crowdsourced real utterances on Amazon Mechanical Turk (AMT).
To do so, we created 75 pairs of charts such that in each pair, the second chart can be made from the first one through up to three editing operations.
We present these pairs to crowdworkers, and ask them to describe in natural language how they would make the edits (\eg \sutt{change the color in the bar chart, and then rescale}). 
We collected 100 descriptions per chart pair, resulting in $\sim$5.4k utterances, and kept 4.7k after manual validation. We configured the AMT HIT such that each crowdworker would describe ten chart pairs. We collected the dataset from $\sim$400 unique workers.
We further augment the dataset with synthetic examples by (1) paraphrasing these utterances through back translation and crowdsourcing~\cite{iyyer2018adversarial}, and (2) creating syntactic templates inspired by Malandrakis \etal~\cite{malandrakis2019controlled}.
The augmentation is a common technique for accelerating early collections of user intents~\cite{malandrakis2019controlled}.
In total, our dataset contains 10.7k utterances \footnote{https://github.com/microsoft/VisTalk}. We use 80\% of the data for model training and the rest for testing.
In the training process, we use the Adam optimizer. The batch size is 32, and the epoch is 150.
Our model worked well on the test set: the operation classification and the tagging F1 scores for objects and parameters were 94.75\%, and 97.34\%, respectively. The sequence labeling F1 score is evaluated at entity-level using seqeval python library\cite{seqeval}.

The performance is promising, and partially benefits from the fact that the data binding and data abstraction in Stage 1 (Section~\ref{sec:abstraction}) eliminates noise in entity recognition. 
However, another primary factor driving the high performance would be our modest coverage of the utterances. The chart pairs we created to collect data involves 1-3 operations, which were easy to be expressed by the crowdworkers and learned by the model. 
It should be noted that the accuracy of the model may decrease when the user utterances become more complex.
In fact, being ``data hungry'' is a significant bottleneck for deep learning models: to reach the best performance, a large amount of training data is required to provide enough training signal on more rare and complex patterns. 
Limited by the training data, we only see our implementation of the deep learning model as a starting point and a proof-of-concept.
We note that to maximize the utility of such models, future work should collect a large number of utterances, and additionally rely on data augmentation techniques as we explored.

\subsection{Stage 3: Editing Action Synthesis}
Given the recognized information, we then map between the independently outputted operations, objects, and parameters, to organize them into a sequence of editing actions.
We take a bottom-up approach to traverse the operation list. 
Because each operation determines its expected parameters and objects, we match the corresponding parameters and objects among the candidates.
For example, \sop{setColor} would only accept colors as parameters. Thus, we traverse the list of parameters and objects to search for an entity \sutt{blue} labeled as \spar{color}.
If there is no parameter or object for an operation, we fill it with a \texttt{*} mark to denote the default state. If there is more than one parameter for an operation, we duplicate the operations into two or more and synthesize more editing actions.
We further heuristically rank these actions based on the category they belong to. We prioritize global operations over local operations. The operations in the category of data operations (\eg sum) and encoding operations (\eg bind a colum to x axis) should be executed before mark operations  (\eg changing the shape of marks), styling operations (\eg changing the color or the title content), layout operations (\eg move upward), and annotate operations (\eg add text labels). The editing actions are then ordered by their appearance in the utterance. 
The operation mappers can then implemented to map the editing action sequence to application-specific operations.
\subsection{Guideline for Reusing the Interpreter}
\revise{
Instead of directly translating NL utterances into application-specific commands, our NL interpreter abstracts out the interpretation of editing intents to enable reusability, following the existing NLI framework \cite{narechania2020nl4dv}.
The only difference across visualization applications is the implementation of operation mapper, as shown in Figure~\ref{fig:authoring_pipeline}.
When users' input is complete and accurate, the interpreter can support easy binding and the editing action can be directly mapped to executable application-wise commands.
When users input application- or context-dependent utterances, we expect the application to clarify the vague and relational parameters we define in Section~\ref{subsec:parameter}.
For example, when users input \sutt{change color to cyan} in Figure \ref{fig:examples}(a), the objects corresponding to the operation \sop{setColor} is underspecified and denoted as ``*'' by the NL interpreter. Therefore, the system should trace back to find the most recent utterances with the objects where the operation can be applied.
Here, we discuss some implications on such handling. 
}

\textbf{Operation mapper.}
The visualization applications should implement an operation mapper that maps the interpreted editing actions into tool-specific operations.
If the operations are clearly specified, the execution engine blends the objects and calls the related functions. If the operation is not supported, the execution engine can return error messages to notice the users. If the operation is not supported but related operations are supported, the application can further show simple examples to help users understand the functions of the system. When the actions are not fully specified, the application could predefine a set of rules to recommend proper actions for under-specified utterances. \revise{Applications could further adopt more advanced chart recommender such as Draco \cite{moritz2018draco} to resolve the underspecified editing actions to improve the editing experience.}

\textbf{Manage Contexts.}
To better understand users' editing intents, the applications could manage users' sessions and maintain a list of objects on canvas with corresponding attributes. If the objects are clearly referred, the application simply looks for the objects and applies operations to the target objects.
For objects with selectors, applications can traverse through the parameters of the objects to find out the referred objects. If the objects cannot be resolved, the application could infer the target objects by examining the recent objects being edited, and recommend possible editing. Alternatively, the application could also return error messages to notify users to clarify their intents.

\textbf{Resolve Ambiguities.}
The applications should build support and disambiguate between the three types of parameters: exact, vague, and relational ones.
For exact parameters, applications can directly map the extracted parameters to the data value and check whether the value can be legally assigned. For example, \spar{10px} for height can be directly assigned to the objects. 
For vague parameters, applications can develop a set of metadata or rules to explain vague parameters supported by the NL interpreter. For example, when users input a vague parameter, the application looks up the parameter \spar{red} and yields the color code RGB(255, 0, 0). 
For relational parameters, applications need to parse the parameters into the changing directions of parameter adjustment. For examples, \sutt{on the top} for \sutt{put it on the top of the \textit{US} bar}, means the object should be placed at a point that has smaller distance to the top border than the US bar, but with the same distance to the left/right border as the \textit{US} bar. Therefore, applications should extract the positions for the \textit{US} bar, and calculate a proper position number for the newly added object. \revise{To recommend the proper position for newly added objects, advanced chart layout algorithms \cite{wu2021learning} can be adopted to suggest system generated design under the constraints of users' intents. Similar design recommendations can also be implemented by the systems to suggest colors, encodings, etc. 
To enable users to conduct even more fine-tuned operations, other modalities such as mouse and touch should be introduced. 

\textbf{Multimodal Interaction.}
Obviously, natural language is not always the best choice of input when conducting visualization editing, and we believe it acts as a complementary input modality to traditional WIMP interaction. 
Other forms of input (e.g., speech, mouse, touch, and pen) could also be combined to support more multimodal interactions.
For example, instead of expecting users to grasp the jargon for describing visualization objects (\eg bar, scatterplot, etc.), they can \emph{select} objects before voicing natural language commands relevant to the selected object.
Furthermore, voice input is also a design choice to combine multiple modalities, where users could save the efforts of typing NL utterances. This could be realized by introducing a speech recognition \cite{riccardi2005active} module at the beginning of our current pipeline to transform voice into text. However, speech-based NLI faces unique challenges (\eg triggering speech input and transcription errors).
The NL input could also be interleaved with input from other modalities. For example, when resolving editing actions that are not partially specified or ambiguous,  visualization systems could also prompt other modalities of user interfaces, such as interactive widgets to help users clarify their intents. Multimodal coreference resolution is an important task as users may input NL queries that follow their direct manipulations on the interface \cite{Kumar2017,Shen2021a}. So the design of operator mapper should involve heuristics to handle coreference resolution in multimodal sense based on the design of the NL interactions of the authoring system.
}


\textbf{Enhance Discoverability.}
Users may not be aware of what operations are available to the system and whether there is a preference for a particular language structure in the system. System discoverability is considered an essential factor that improves the user experience \cite{srinivasan2019discovering,Setlur2020}. It could happen that the interpreter could parse the utterances into editing actions, but the system is not able to execute the operations. Another possibility is the system has the corresponding operations but the interpreter could not understand the utterance correctly.
For the former situation, the system could explain its scope of capability based on the editing actions. For the latter situation, the system could either notice the users to take other modalities to complete their tasks, or educate users on how to phrase queries that can be interpreted correctly by the system. In terms of interaction design, text autocompletion can be leveraged to help users precisely complete NL input; interactive widgets with data/visualization previews can be useful for visualization authoring to enhance discoverability.

\begin{figure}[tb]
    \centering
    \includegraphics[width=\linewidth]{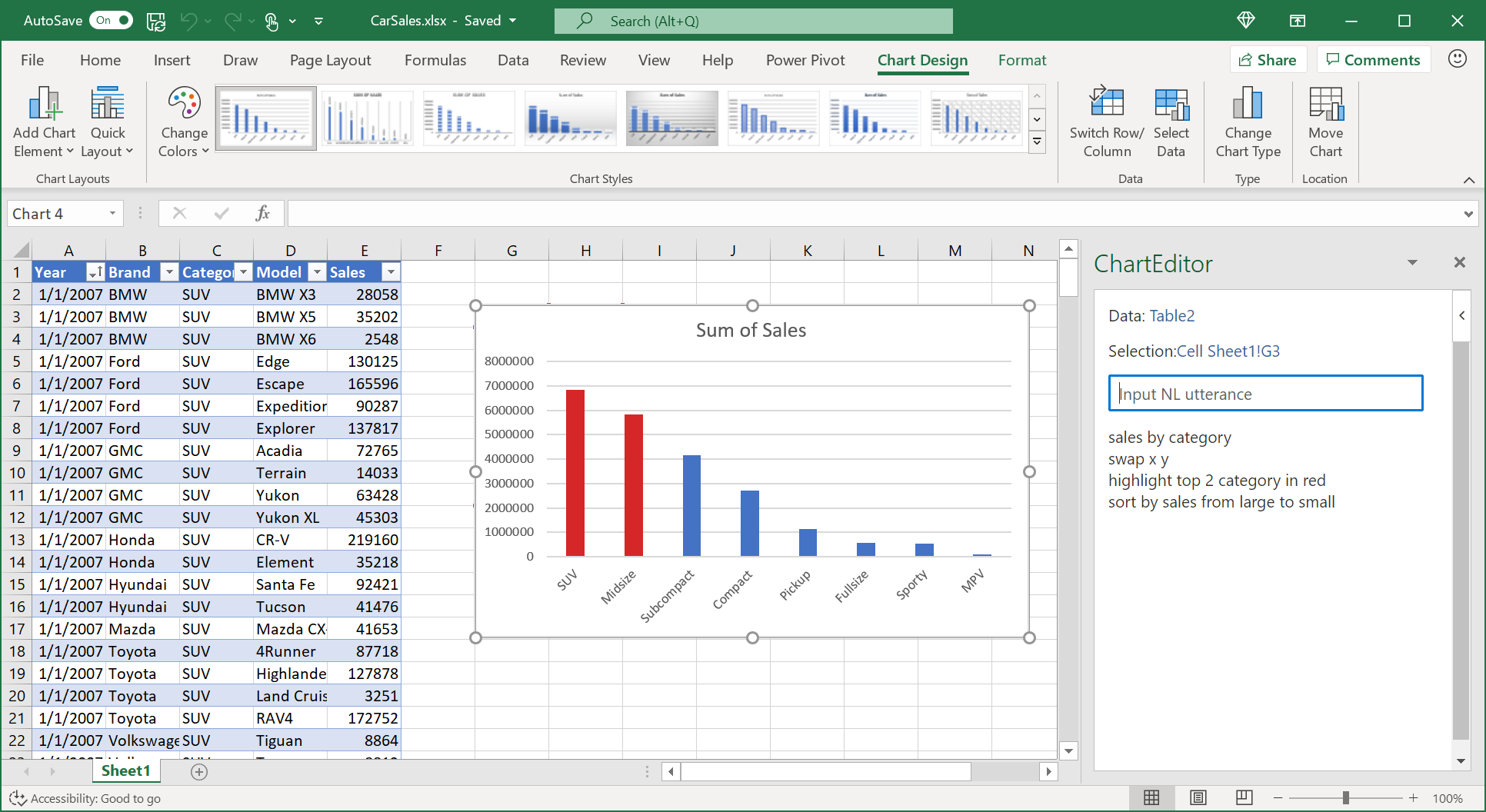}
    \caption{An Excel add-in for chart creation and editing. Users can input NL utterances in the input box. The selected chart updates accordingly.}
    \label{fig:add-in}
    \vspace{-3px}
\end{figure}
\begin{figure*}[tb]
		\setlength{\abovecaptionskip}{3pt}
		\setlength{\belowcaptionskip}{3pt}
    \centering
    \includegraphics[width=0.95\linewidth]{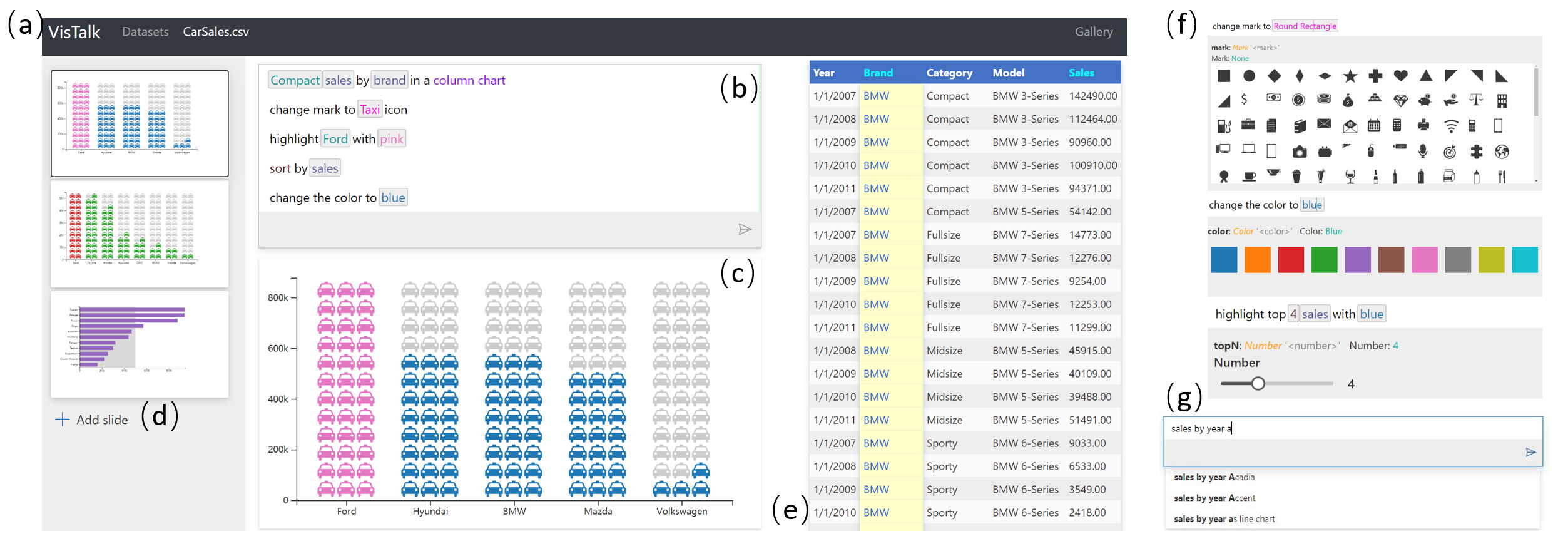}
    \caption{VisTalk interface (a). Users can type in their natural language utterances in the text box (b) to customize their visualization design (c). Users can author different charts at the same time (d). The table view shows the underlying dataset (e). 
    Users can click the words as interactive widgets to resolve ambiguities (f). The auto-completion panel pops out during the input process (g).
    }
    \label{fig:interface}
     \vspace{-3px}
\end{figure*}

\section{Example Applications}
\label{sec:application}

As proof-of-concept, we build two example applications 
of NL-based visualization editing system powered by the NL interpreter in Section~\ref{sec:interpreter}. 
The systems are example implementations of the \emph{application} component in Figure~\ref{fig:pipeline}.

\subsection{Excel Chart Editor}
\revise{
Based on our NL interpreter, we build Excel Chart Editor, an Excel add-in to integrate NL-based chart editing into Excel. 
As shown in \autoref{fig:add-in}, we provide a natural language input panel for users to type in their natural language commands. 
The chart editor creates a new chart when users type in their first sentence and continuously update the chart as they input follow up utterances.
Since Excel already offers rich functions of creating and editing charts on canvas, the primary implementation effort is on mapping the editing actions to the predefined executions\cite{excel}.
This includes resolving object selectors, executing data queries, and converting vague parameters into accurate default values. 
For example, the editing action \sact{setColor}{mark}{color=red} can directly map to \texttt{setSolidColor(color)} method in \texttt{Excel.ChartFill} interfaces.
Further, we take a set of simple heuristics to resolve ambiguity. For example, for input utterance \sutt{sort}, users do not specify the order and the data field. We by default sort the fields corresponding to the y-axis in descending order. When user input \sutt{make the line stroke wider}, we increase the line stroke width by 50\%.

Note that the implementation of the Excel add-in does not concern about the textual inputs, they are only parsed by the NL interpreter. 
It demonstrates how our NL interpreter and editing actions can be plugged into existing tools, and seamlessly augment the larger pipeline of data analysis and presentation workflow.

}

\subsection{VisTalk}\label{sec:vistalk}

We also develop VisTalk, an NL-based standalone chart creation tool.
Similar to the Excel add-in, VisTalk takes users' natural language input through a simple text input box (Figure~\ref{fig:interface}(b)), and automatically re-renders the visualizations to show changes accordingly (Figure~\ref{fig:interface}(c)). 
But uniquely, while Excel add-in concerns augmenting existing tools, we use VisTalk to show how developers might design their own applications while maximizing the utility of our NL interpreter. 

In particular, VisTalk demonstrates how applications can resolve ambiguities in the identified editing actions.
As mentioned in Section~\ref{sec:evaluation}, users commonly submit queries that have missing information. 
Compared to the examples in Figure~\ref{fig:taxonomy} (\sutt{give me a bar chart of Sales by Brand in 2012}), an utterance \sutt{Sales by year} misses the \sop{setChatType} operation, and can only be parsed into
\sact{BindX}{x axis}{data field="Year"} and 
\sact{BindY}{y axis}{data field="Sales"}.
In response, VisTalk recommends chart types from partial specifications of visual charts, following the research of chart recommendations \cite{setlur2019inferencing, mackinlay2007show}, 
In the above example, VisTalk would recommend a bar chart based on the data types.
VisTalk also adopts interactive widgets (Figure \ref{fig:interface}(f)) and utterance auto-completion (Figure \ref{fig:interface}(g)) to disambiguate parameters (\eg for elemments). 
\revise{
For example, users can click on the keywords within the input utterance to specify and refine the choice of icons, colors, chart types, and values in a pop-up window (Figure \ref{fig:interface}(f)).}

\begin{figure}[t]
    \centering
    \includegraphics[width=\linewidth]{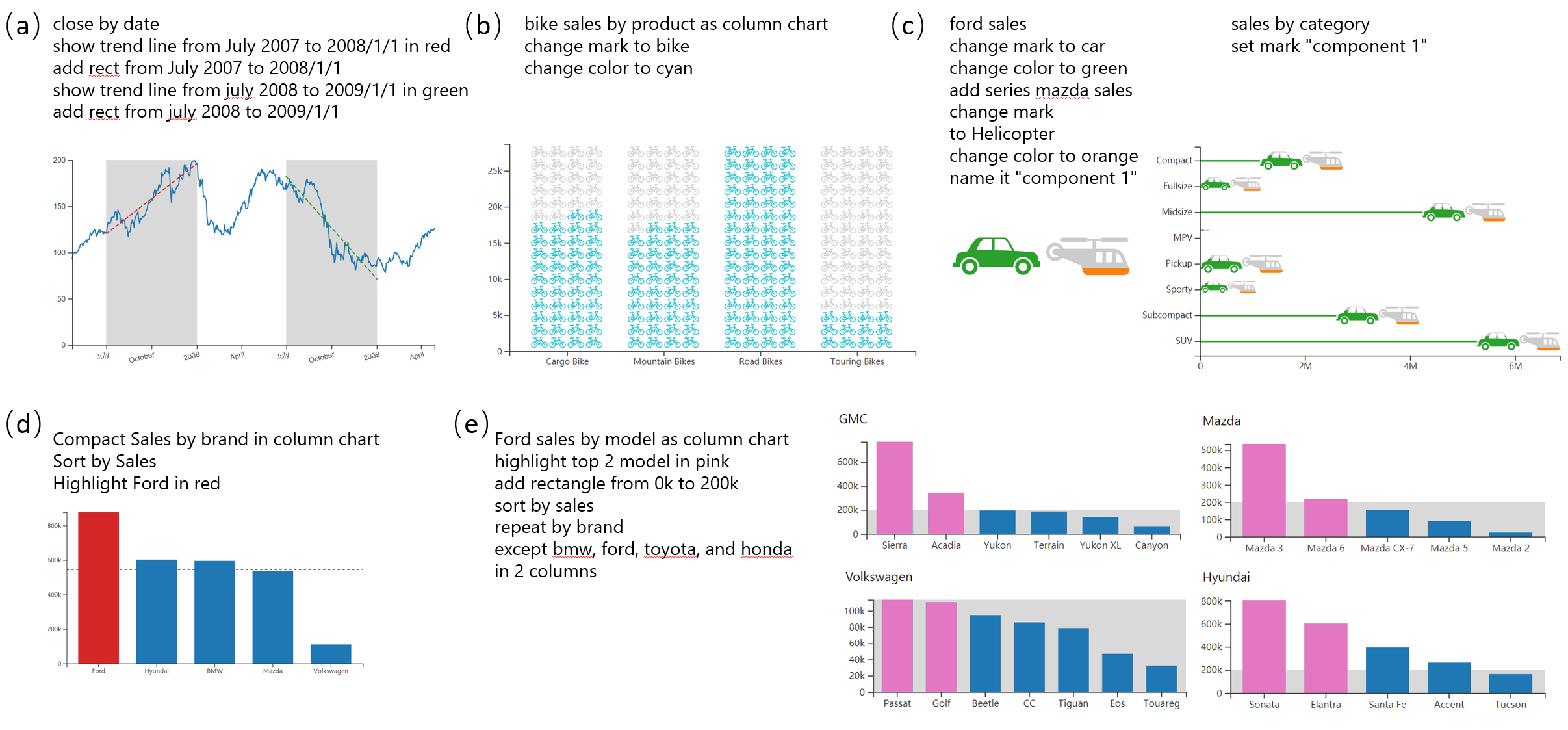}
    \caption{Examples wit VisTalk. (a) is based on a product sales dataset, (b) and (e) are based on a car sales dataset, (c) is based on a shark attacks dataset, and (d) is based on a stock datatset. The terms displayed in bold and italic are column names and values in data, respectively. 
    }
    \label{fig:examples}
     \vspace{-12px}
\end{figure}

\textbf{Example Gallery:}
To demonstrate the usability of VisTalk, we produce a variety of charts with example natural language utterances that specify the visualizations as shown in Figure \ref{fig:examples}. 
These visualizations cover a wide range of operations that users may take in real world visualization authoring scenarios, and show that VisTalk can enable the creation of basic charts with simple operations such as data operations, encoding operations, and annotate operations with NL utterances.

\subsection{Exploratory Study}
\label{sec:evaluation}
With VisTalk at hand, we further conduct a user study to understand how users interact with visualization systems to construct visualizations through natural language.
\subsubsection{Participants}
 We recruited 12 participants (three females and nine males, age ranging from 22 to 45) who had normal or corrected-to-normal vision. The participants include undergraduate and graduate students, data analysts, researchers, and software engineers. 
 They are general users who need to create visualizations to present data in daily work. All of the participants had used Excel to create charts before, while five of them had used Tableau or Power BI to create charts. 

\subsubsection{Study Procedure}
Our user study lasted for about one hour.
We first surveyed participants' background information with a questionnaire.
Next, we provided a tutorial outlining the features of VisTalk with two examples.
As a warm-up exercise, we encouraged them to freely explore the tool with one sample dataset.
Then, we provided the participants with \textbf{five visualization reproduction tasks}. \revise{The participants are asked to reproduce the charts that we provided, ordered by complexity.}
These tasks included the reconstruction of five charts: (1) one basic bar chart, where participants only need to specify data and encoding; (2) two charts with annotations (i.e., trendline, average line, and annotation band); (3) one pictograph, where participants need to modify the shape of the marks; and (4) one chart with multiple views, where participants need to repeat the chart design on a data column, similar to Figure~\ref{fig:examples}(e).
For each task, we displayed the target chart, described the underlying datasets, and asked participants to reproduce the chart in VisTalk using natural language utterances.
Afterwards, participants rated their experience with VisTalk in the form of a five-point Likert scale~\cite{likert1932technique}.
We further collected their free-form feedback through a semi-structured interview.

\subsubsection{Results} 
\begin{figure}[!tbp]
    \centering
    \includegraphics[width=0.95\linewidth]{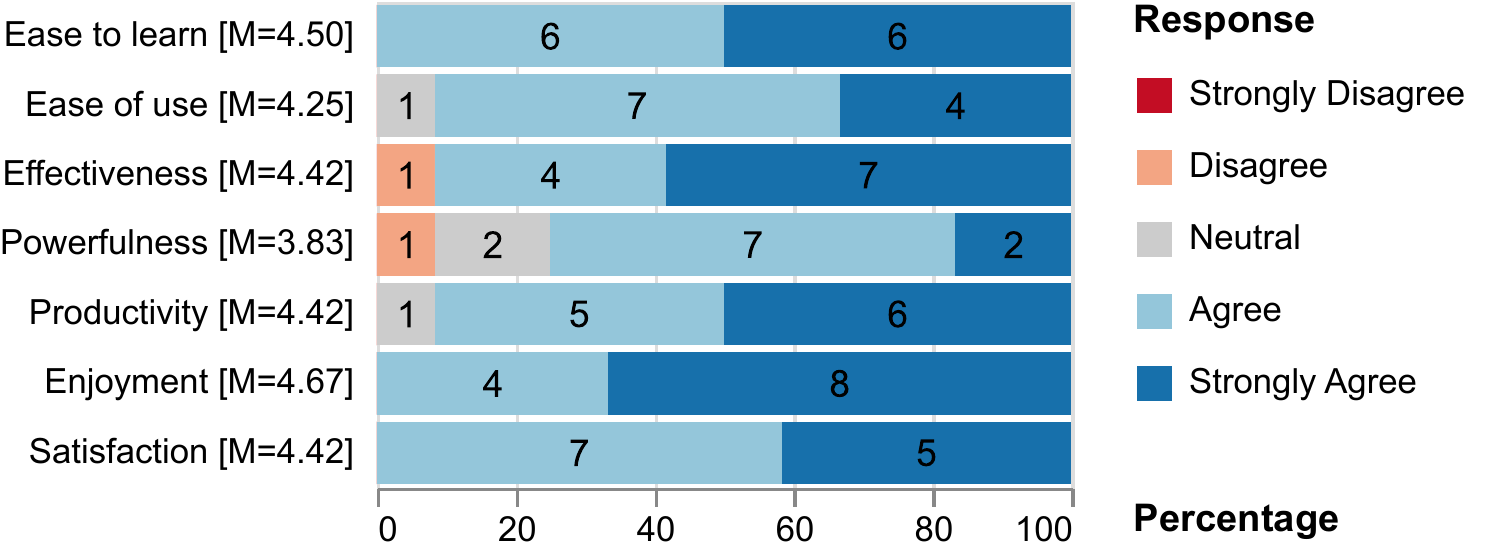}
    \caption{User ratings of VisTalk system with a 5-point Likert scale.}
    \label{fig:time}
    \vspace{-12px}
\end{figure}

\textbf{Subjective Satisfaction:}
Overall, the participants are positive about the system. Users highly rated the experience of using VisTalk (M = 4.36, SD = 0.69). Figure~\ref{fig:time} lists the feedback of VisTalk. Eleven of twelve users agree VisTalk is easy to learn and easy to use. Eleven users think VisTalk has improved their productivity. Twelve users felt satisfied. The feedback also reveals space of improvement. Our participants rated it relatively low in terms of the powerfulness of the system and the effectiveness of completing their jobs. It may be helpful to design more functions and interactions for VisTalk for future work.

\textbf{Result Analysis:}
All the participants have completed the tasks. The participants learned quickly after the training. While they were typing, the system were parsing the query automatically in the meantime. If the utterances were successfully parsed,  specific parameters in the utterances were highlighted (Figure \ref{fig:interface} (b)), and the visualization was updated. If the system could not parse it, the user could iteratively modify the queries to make them correct.

To understand the complexity of utterances, we use mean length of utterances (MLU) \cite{oviatt1997mulitmodal}.
The number of words used ranged from 1 to 11.
The average MLU across 12 users is 3.61 words per utterance (SD=1.26).
On average, the participants used 2.4 utterances to complete a visualization creation task (SD=0.72).
The average time of creating one chart was 139s.
 We found the participants tend to avoid using long sentences, but instead, shorter utterances for incremental configurations. Since the input textbox in VisTalk is directly editable, users tend to correct the NLs inplace instead of appending additional utterances.

Our participants had a positive experience authoring with natural language.
For simple tasks, users usually only type in some keywords to see how the systems can parse their intentions. For example, after one user (P2) tried ``trend'' and then ``add trend'', the system automatically added a trend line to the line chart. The participant said, ``\textit{It worked well. Natural language makes the thing easy.}''
Participants also mentioned that it is easy to accomplish complex tasks with natural language, because it allows them to combine many operations within a simple sentence in the way that meets their needs. One user (P3) commented, ``\textit{The experience is very good! In the past I need to search the menus and click the right buttons to perform the tasks. Sometimes it costs a long time to find the right menu item. Now I don't need to think of where they are.}'' 

The majority of participants (10/12) mentioned that the natural language interface saves their time and efforts. Most participants (8/12) also said they will use natural language as their first choice if their familiar systems support similar functions. P9 said, ``\textit{To design a simple chart, using one simple sentence is enough. It is completely different from using mouse to select menus, select the data field from the dataset, specify chart types, and do further adjustments.}'' Impressed by the parsing capability of VisTalk, P1 mentioned ``\textit{It is surprising that the system could perform correctly even when the sentence was very short and I misspelled some words and had grammatical errors.}''

The participants (12/12) find the system interactions easy to learn, even for those without much experience in visualization. One participants stated, ``\textit{I can quickly understand the effects of the commands.}'' One of the participants who was not familiar with visualization creation tools said, ``\textit{Now I don't have motivations to learn to use traditional visualization tools. I will come to VisTalk if I need to create charts. It seems much easier to learn.}'' P3 also mentioned, ``\textit{It is friendly to non-expert like me. I don't find difficulties of learning to use this tool. It gives me prompt feedback while I am typing and I feel I can complete the tasks easily. The overall experience is quite smooth.}'' 

Although the chart recommendation functions took simple strategies, many of our participants (7/12) mentioned the system recommendation is helpful. One participant said, ``\textit{The default charts were well-designed and I don't think I need to do further modifications.}'' Three  participants mentioned they enjoyed using short phrases to interact with systems' default recommendations. P9 said, ``\textit{I am using these keywords to explore this system, similar to the experience of using search engines. I enjoy the way this system gives me surprises.}''


Participants also found difficulties when using NLIs. 
For example, when a user (P11) wanted to change the color of the charts, he typed \sutt{highlight in blue} and the system did not correctly execute the command because no object to highlight is provided.
Some common errors in NLIs also happen in the study, such as synonyms (e.g., ``bike" and ``bicycle"), ambiguous utterances, and spelling errors. For example, one user (P2) said, \sutt{set bike mark to bike}. The system can not parse the query as ``bike" is ambiguous (the first is mark and the second is value). One user (P5) also typed some specific configurations (e.g., \sutt{arrange by 4 × 2} can not be parsed but \sutt{arrange in 2 columns} works) and system control (e.g., \sutt{refresh view} and \sutt{re-center graph} ask the system to load and position the visualizations) that are not supported by the interpreter. Some analysis-oriented NLI users may pose queries that expresses an intent of data analysis, instead of chart editing (e.g., \sutt{What is the relationship between sales and product?}).
Some other participants also felt they couldn't express their requirements especially at the beginning of the tasks. ``\textit{Sometimes I forgot the words and felt hard to describe my requirements.}'' One user applied to open and read the examples in the tutorial again at the start of the tasks. Another user opened search engine to look for a correct word. 
Although all of them got used to the natural language experience after a while, we find that cold start can be a problem for users that are not familiar with the system. The reason may be that users are dim about how much the system could tolerate their vague or inaccurate expressions. To solve this problem, we believe more sufficient guidance and feedback can address this issue. The interface should also provide recommendations to give hints about potential choices that they can give commands.

\section{Discussion}
By taking an one-interpreter-for-all schema, we not only save the efforts of supporting NLIs for visualization authoring applications, but also alleviate the need for users to get familiar with specific concept models.
Still, with NLIs freeing users' mental models from any specific design paradigms~\cite{satyanarayan2019critical}, the overflexibility can lead to some challenges in practice. 
Below, we discuss three gaps between what people would say and how a system might (incorrectly) respond, all emerged from NL commands being overly flexible, and discuss potential future work.

\revise{ \textbf{Gap 1: Users express commands in diverse patterns, but NL parsers cannot recognize them all.}}
Users that have different backgrounds, or are familiar with different visualization tools, can express the same objective in drastically different ways.
For example, Section~\ref{sec:actions} enumerates the different visualization creation types; to create the chart from scratch, the utterances can be as high level as just mentioning the data columns (and the system is expected to automatically infer the chart type), and it can also go as low-level as binding each specific mark channel (\eg rectangles) individually.
To achieve the seamless switch between these types, the \emph{interpreters} should be able to understand different commands. 
Following the deep learning approach as in our interpreter (or, even to generalize it with recent pretrained NLP models~\cite{Devlin2019}), a crucial future step would be to collect a diverse set of training utterances that express the same intent in various usage scenarios and levels of details.
This can be achieved by feeding crowd labelers with richer \emph{visualization objectives}. 
We can even diversify utterances by constraining them on \emph{construction workflows}, \ie to ask expert users to write utterances that can only be parsed into a serial of preset actions.
Moreover, it is possible to collect naturally occurring utterances without any predetermined objectives or application workflows.
Recently, researchers have asked users to freely submit any possible queries to analyze charts, so to build taxonomies on representative utterances~\cite{srinivasan2021collecting}, and synthesized natural language to visualization (NL2VIS) benchmarks from NL2SQL benchmarks\cite{Luo2021}.
Though still targeting at visual analysis, these work shed light on possible designs to collect diverse utterances. We can then further augment the datasets by paraphrasing these commands.

\textbf{Gap 2: Users have various editing objectives, but the downstream application may not implement them all.} While the capability of parsing natural language utterances increases, our applications might be \emph{designed} to only focus on a subset of visualization creation and an editing actions. For example, how should a system react to \sutt{change the rectangle to circle}, if it only wants to support analytical interactions, and therefore does not allow mark customization? Would it disappoint people, if by design an application ignores more utterances than another tool, even when the interpreter parses them correctly? User studies on how users react to the boundary of applications are interesting. Alternatively, we can improve command recommendation, make suggestions to users when their utterance does not belong to the supported function type, and help people understand the scope of application functions.

\textbf{Gap 3: Users use NL for various intentions simultaneously, but the framework does not go beyond authoring.} 
Along with our work, there exist various frameworks that tackle analysis, authoring, data processing, etc. separately.
However, users may interchangeably express these needs through natural language all at once, and it is impractical to expect users to swiftly switch between these tools.
Here, integrating analysis-oriented and authoring-oriented NLI seems promising.
Just as we have hinted in Section~\ref{sec:intro}, users can start with the visualizations produced from analysis-oriented NLIs and make further configurations on top to improve visual interaction with authoring-oriented NLIs. 
Furthermore, recent NL2SQL~\cite{Azcan2020,Affolter2019} advances can also be utilized for low-level data-driven queries. 
One challenge could be to design additional query type classifier modules that can identify analysis-oriented, authoring-oriented, and database-oriented queries and, more importantly, how these operations should be correctly ranked so that the final visualization reflects all the requirements correctly.


\section{Conclusion}
In this paper, we explore a natural language-based visualization authoring pipeline, which supports the understanding of visualization construction commands. We propose the definition of editing action that describes users' visualization editing intents, defined as tuples of operations, objects, and parameters. The editing actions bridge the gap between the NL interpreter and visualization applications. We further implement a deep learning-based NL interpreter, to extract operations, objects, and parameters from users' utterances. From these extracted information, we synthesize the editing actions for visualization applications to handle. Based on the NL interpreter, we demonstrate the utility of our pipeline and NL interpreter with two example applications, an Excel chart editor and a proof-of-concept authoring tool, VisTalk. 
To assess our approach, we further conduct a user study with VisTalk to understand how users edit visualizations through natural language. Our study is a first step towards NL-based visualization authoring.

\bibliographystyle{abbrv-doi}

\bibliography{ref}

\clearpage
\end{document}